\documentclass[final,5p,times,twocolumn,authoryear]{elsarticle}
\biboptions{authoryear}
\journal{Molecular Astrophysics}

\begin{document}
\begin{frontmatter}

\title{An optical spectrum of a large isolated gas-phase PAH cation: C$_{78}$H$_{26}$$^+$}

\author{Junfeng Zhen$^{1,2,*}$, Giacomo Mulas$^{2,3}$, Anthony Bonnamy$^{1,2}$, Christine Joblin$^{1,2}$}

\address{$^{1}$Universit\'e de Toulouse, UPS-OMP, IRAP, Toulouse, France}
\address{$^{2}$CNRS, IRAP, 9 Av. colonel Roche, BP 44346, 31028, Toulouse Cedex 4, France}
\address{$^{3}$Istituto Nazionale di Astrofisica - Osservatorio Astronomico di Cagliari, via della Scienza 5, 09047 Selargius (CA), Italy}
\cortext[$^{$*$}$]{Correspondence author: Junfeng Zhen; Electronic mail: junfeng.zhen@irap.omp.eu }
\begin{abstract}
A gas-phase optical spectrum of a large polycyclic aromatic hydrocarbon (PAH) cation - C$_{78}$H$_{26}$$^+$- in the 410-610\,nm range is presented. This large all-benzenoid PAH should be large enough to be stable with respect to photodissociation in the harsh conditions prevailing in the interstellar medium (ISM). The spectrum is obtained via multi-photon dissociation (MPD) spectroscopy of cationic C$_{78}$H$_{26}$ stored in the Fourier Transform Ion Cyclotron Resonance (FT-ICR) cell using the radiation from a mid-band optical parametric oscillator (OPO) laser. 

The experimental spectrum shows two main absorption peaks at 431\,nm and 516\,nm, in good agreement with a theoretical spectrum computed via time-dependent density functional theory (TD-DFT). DFT calculations indicate that the equilibrium geometry, with the absolute minimum energy, is of lowered, nonplanar C$_2$ symmetry instead of the more symmetric planar D$_{2h}$ symmetry that is usually the minimum for similar PAHs of smaller size. This kind of slightly broken symmetry could produce some of the fine structure observed in some diffuse interstellar bands (DIBs). It can also favor the folding of C$_{78}$H$_{26}$$^+$ fragments and ultimately the formation of fullerenes.

This study opens up the possibility to identify the most promising candidates for DIBs amongst large cationic PAHs.

\end{abstract}


\end{frontmatter}

\section{Introduction}
\label{sec:intro}
The aromatic infrared bands (AIBs), strong emission features at 3.3, 6.2, 7.7, 8.6, 11.2~$\mu$m, dominating the near and mid infrared (IR) spectrum of many interstellar sources, are generally attributed to IR fluorescence of large ( $\sim$ 50 C atom) PAH molecules pumped by UV photons \citep{all89,pug89,sel84,gen98,tie13}. These PAH species are found to be ubiquitous and abundant, containing $\sim$ 10\% of the elemental carbon. They are expected to play an important role in the ionization and energy balance of the ISM of galaxies \citep[and references therein]{tie08}. They have also been proposed as an important catalyst for the formation of molecular H$_2$ in photodissociation regions \citep{bos15}. In addition, PAHs are also considered as potential carriers for the DIBs \citep{leg85,cra85,sal99}. The leading idea in this proposal is that if PAHs are as abundant as they need to be to produce the observed AIBs, and since PAHs have prominent bands in the visible when either ionized or large enough, such bands must contribute to DIBs. The main arguments for PAHs as likely DIB carriers were reviewed relatively recently by \citet{cox2011}, and a critical review of the current status of research in this direction is available in \citet{salama2013}. 

Over the years, many experimental, theoretical and observational studies have investigated the spectroscopic properties of PAHs in relation with DIBs, ranging from the smallest species \citep{salama1992} to relatively large ones \citep[see e.g.][]{weisman2003,huisken2013} and to PAH derivatives \citep[see e.g.][]{hammonds2009,rouille2013}. Some effort was spent in trying to study large numbers of PAHs systematically, attempting to single out general trends that would hopefully allow to choose the most promising candidate DIB carriers in the vast population of this chemical family \citep[see e.g.][]{ruiterkamp2005,weisman2003,tan2009}. It is thus now understood that the onset of absorption due to electronic transitions tends to shift redwards with the increasing size of PAHs, and that very fast non\textendash radiative transitions severely affect the spectral shape of PAH bands, due to lifetime broadening effects \citep{pino2011}. This latter lifetime broadening effect tends to be much stronger for ions than for neutrals, with the former thereby exhibiting shallow, featureless bands, whereas the latter produce bands with some recognisable rotational envelopes. Notably there are DIBs qualitatively matching both kinds of spectral behaviours \citep[see e.g.][]{sarre2013}

Recently four DIBs were identified as due to the fullerene cation C$_{60}$$^+$  \citep{camp15, wal15}, which confirms that large carbonaceous molecules are good carrier candidates for the DIBs. However so far no DIBs could be identified as arising from electronic transitions in PAHs. The reason might be that spectroscopic measurements on PAH cations have concerned  relatively small species, whereas chemical models predict that only large PAHs can survive the UV radiation field even in the diffuse interstellar medium where this field is quite diluted  \citep{lep01,mon13}. \citet{berne12} have shown that closer to stars even large PAHs are expected to be efficiently photodissociated and this photoprocessing could be at the origin of the formation of C$_{60}$ in these regions \citep{berne15}. 

The very large PAH, C$_{78}$H$_{26}$ cation, has been studied in \citep{zhen2014} as a possible precursor of fullerenes. It was selected because its armchair edges give it greater stability than PAHs with zigzag edges \citep{poa07, kos08}, thereby favoring its presence in space \citep{can14}. But due to the limitations of their experimental set-up, \citet{zhen2014} had difficulties to observe its photo-fragmentation behavior, especially the dehydrogenation. 

We here describe our new study of the photo-fragmentation behavior, and a gas-phase spectrum of cationic C$_{78}$H$_{26}$ that we obtained in the range of 410-610\,nm. The spectrum of this large PAH cation is compared with TD-DFT calculations to investigate its electronic structure. The experimental methods and results are described in section~\ref{sec:exp}, the computational methods and results are shown in section~\ref{sec:com}. Section~\ref{sec:dis} compares the experimental and theoretical results, and finally section~\ref{sec:concl} summarizes the main results.

\begin{figure*}[t]
  \centering
  \includegraphics[width=\textwidth]{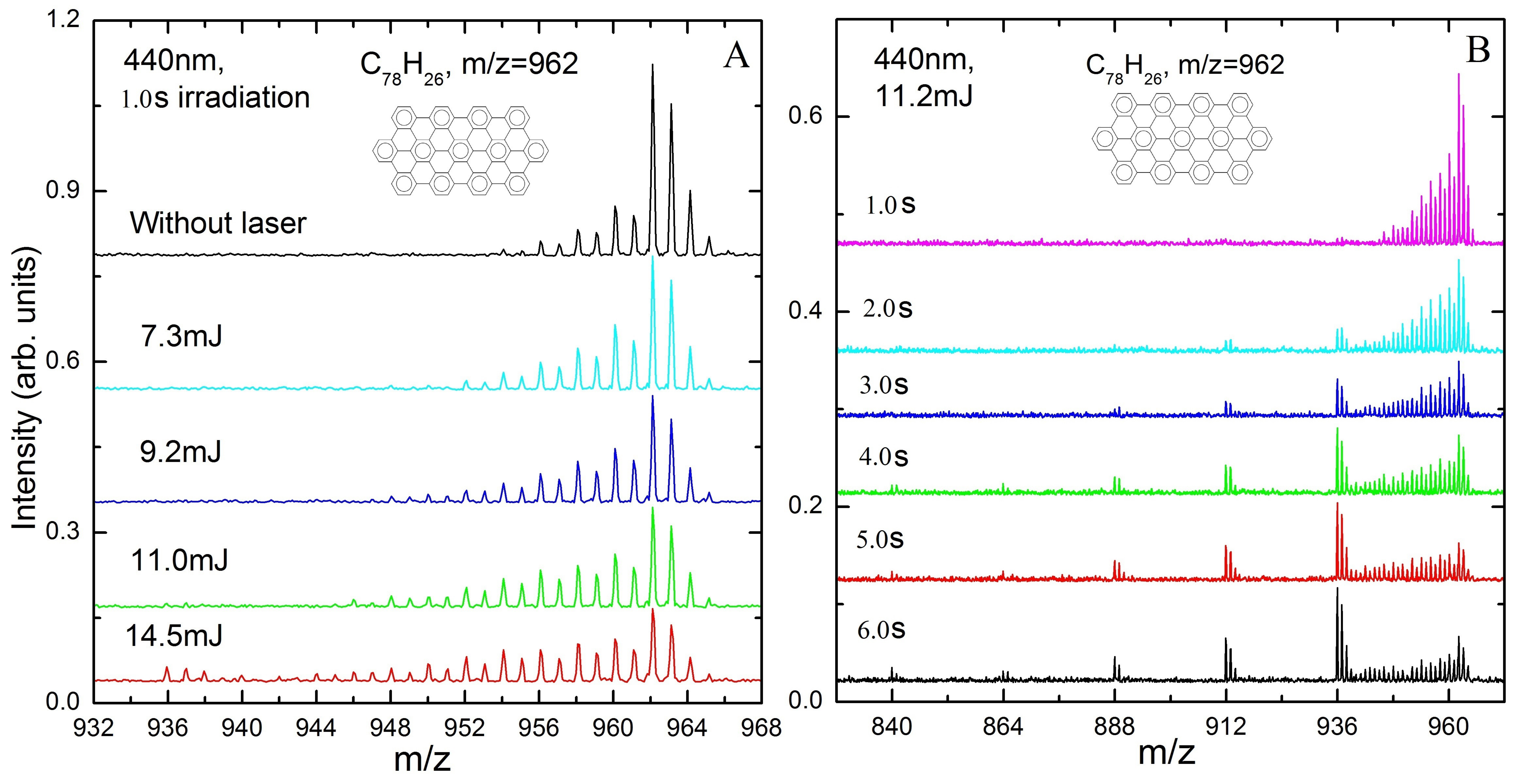}
  \caption{Mass spectrum of the photo-products resulting from irradiation of C$_{78}$H$_{26}$ cation at 440nm: (A) the irradiation time is 1.0 second with different laser intensities; (B) the laser intensity is 11.2 mJ per pulse with different irradiation times.
  }
  \label{fig1}
\end{figure*}

\section{Experimental Methods and results}
\label{sec:exp}
We have studied the C$_{78}$H$_{26}$ cation using PIRENEA ({\it Pi\`ege \`a Ions pour la Recherche et l'Etude de Nouvelles Especes Astrochimiques}), which is an original home-built experimental set-up conceived to perform photo-physical and chemical studies on large molecules and nano-sized particles of astrophysical interest in isolation conditions approaching those of interstellar space in terms of collisions and interactions \citep{job02}.
One of its advantages is that it can be used to study individual photofragments  (which can be an extremely reactive radicals), and very effectively isolate them by expelling everything else from the trap \citep{use10,kokkin2013,wes14}. PIRENEA can be used to study electronic transitions via multi-photon dissociation (MPD) action spectroscopy.
Strengths and drawbacks of this technique are described in detail in \citet{use10}. 

Desorption and ionization of PAH molecules result from laser irradiation of a solid sample of C$_{78}$H$_{26}$ by the fourth harmonic (266\,nm) of a Nd:YAG laser (Minilite II, Continuum).  The sample was produced by the M\"ullen group following the synthesis method described in \citet{muller1997}, which ensures that only one isomer is produced. 
Trapping of ionized species is achieved through the conjugated action of an axial magnetic field of 5 T, generated by a superconducting magnet, and an electrostatic field. The magnetic field confines the ions in the radial direction while the static electric field traps them in the axial direction. The ion trap can be operated at very low temperature ($\sim$ 35\,K) and in ultra-high vacuum ($<$ 10$^{-10}$ mbar), to approach the isolation conditions of the interstellar medium. 
The ions of interest can then be selected and isolated in the ICR cell by ejection of all the other species that could be produced. This is achieved by generating a suitable RF oscillating field containing an appropriately defined spectrum of frequencies, excluding frequencies that can excite the species one wants to retain. 

A mid-band tunable OPO laser system (Panther EX, Continuum; 5 cm$^{-1}$ bandwidth, 5 ns pulse duration) operating at a pulse frequency of 10 Hz is used to irradiate the trapped ions. {\it In-situ} non-destructive Fourier Transform ICR mass spectroscopy is used to analysis the effects of irradiation. Scanning the laser over a grid of wavelengths an action photo\textendash absorption spectrum is obtained \citep{use10}.

In the conditions of the experiment, the photo-processing of PAH cations is heavily biased towards dissociation through the lowest energy channel. Each fragmentation event is initiated by absorption of multiple photons, the exact number depending on the wavelength of the OPO laser and the dissociation thresholds. The absorption of a photon in an electronic transition, followed by fast Internal Conversion (IC), leaves the molecule in a highly excited vibrational state in the ground electronic state, typically on a timescale of the order of 50 fs \citep{mar15}. Intramolecular vibrational redistribution (IVR) then quickly spreads the excess energy among all available vibrational modes, leaving the cation amenable to further excitation through the same sequence of electronic excitation and radiationless relaxation. Whenever the laser wavelength is in resonance with an electronic transition of the cation, the sequential absorption of multiple photons can take place and the dissociation can proceed. For reasons discussed later in Sect.~\ref{sec:dis}, the width of these resonances is very broad due to the experimental technique.

Figure 1 summarizes the photo-fragmentation results of C$_{78}$H$_{26}$ cation irradiation at 440\,nm. The mass spectrum before irradiation of C$_{78}$H$_{26}$ cation reveals some residual fragmentation (peaks due to H loss) as a byproduct of the UV laser desorption and ionization process as well as the presence of the isotopic species containing one to three $^{13}$C at higher mass.
Fig. 1(A) shows the resulting mass spectra of trapped C$_{78}$H$_{26}$ cations upon 440 nm irradiation at different laser intensities; all experiments are performed under the same conditions with an irradiation time of 2.0 s. 
A wide range of dehydrogenated fragment ions are evident in these mass spectra that are attributed to multiple, sequential photon absorption and resulting photo-fragmentation events.  The photo-dissociation of the C$_{78}$H$_{26}$ cation mainly follows sequential 2H (or H$_2$) separations until full dehydrogentaion, leading to the predominance of even-mass fragments C$_{78}$H$_{2n}$$^+$ with n = [0 $-$ 12]. Fig. 1(B) shows the resulting mass spectra of C$_{78}$H$_{26}$ cation upon 440\,nm irradiation with different irradiation times from 1.0\,s to 6.0\,s, at a fixed energy of 11.2\,mJ per pulse. This shows the dehydrogenation photo-products as well as further fragmentation of the bare carbon cluster C$_{78}^+$ leading to  C$_{78-2n}$$^+$ with n = [1 $-$ 4], similarly to what was observed by \citet{zhen2014}.

In Fig. 2, the MPD spectrum of C$_{78}$H$_{26}$ cation is shown. The resulting spectrum was obtained with laser intensity fixed at (7.0 $\pm$ 1.0) mJ per pulse and 2.0 s laser irradiation. Laser wavelengths were scanned in steps of 3~nm. The strongest transitions in the region studied were found to lie at 431\,nm and 516\,nm (FWHM $\sim$ 20nm in the MPD spectrum). The large bandwidth could be related to the lifetime broadening but is more likely related to the high rotational and vibrational temperatures. Sequential transitions after the first one arise from more and more highly excited ro-vibrational levels of the electronic ground state, populated by IC-IVR that redistributes the electronic excitation energy of the previously absorbed photons. 

\begin{figure}[t]
  \centering
  \includegraphics[width=\columnwidth]{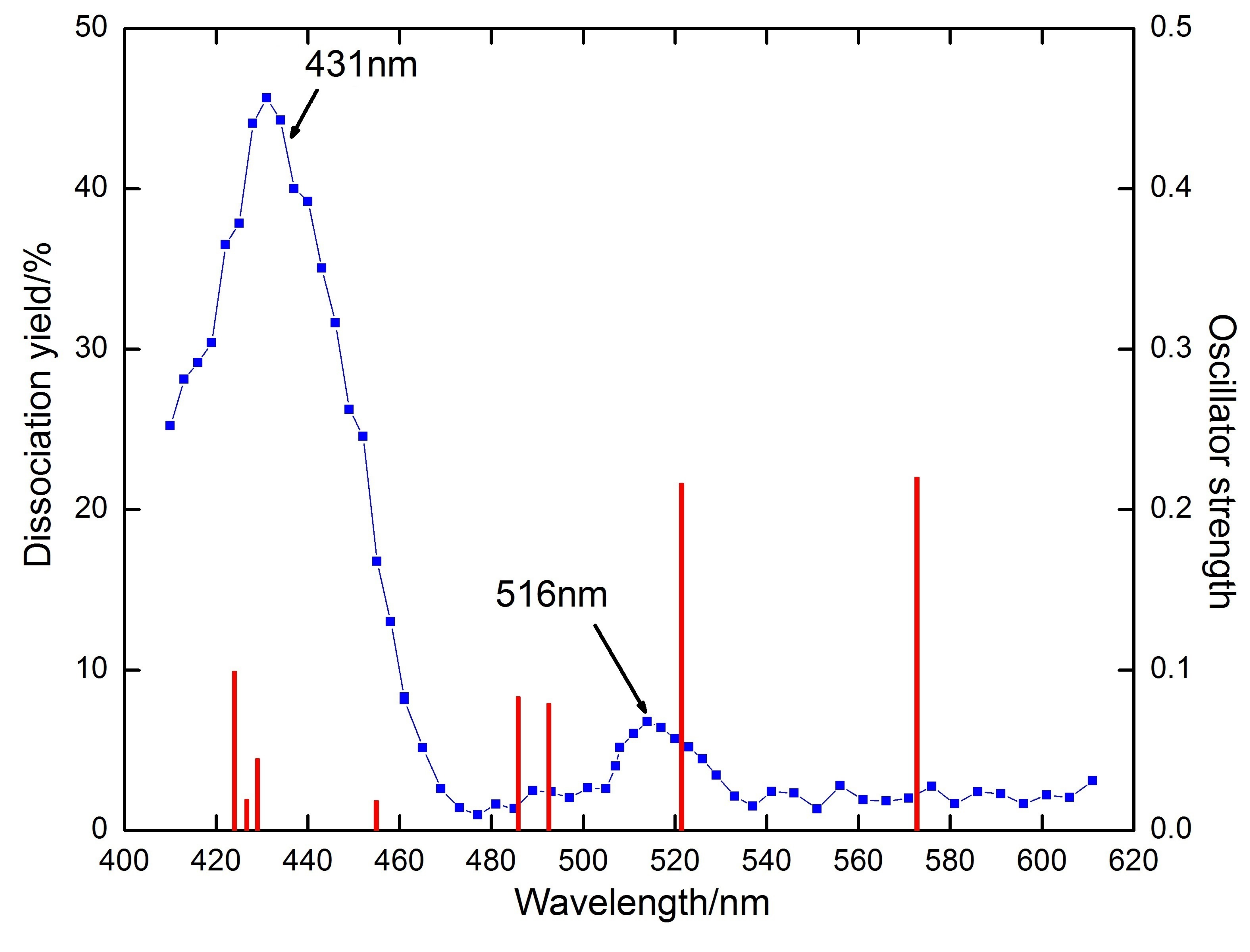}
  \caption{ Measured multiple photon dissociation spectrum of gas-phase C$_{78}$H$_{26}$ cation. Computed transitions are represented by vertical bars.
  }
  \label{fig2}
\end{figure}

\begin{figure}[t]
  \centering
  \includegraphics[width=\columnwidth]{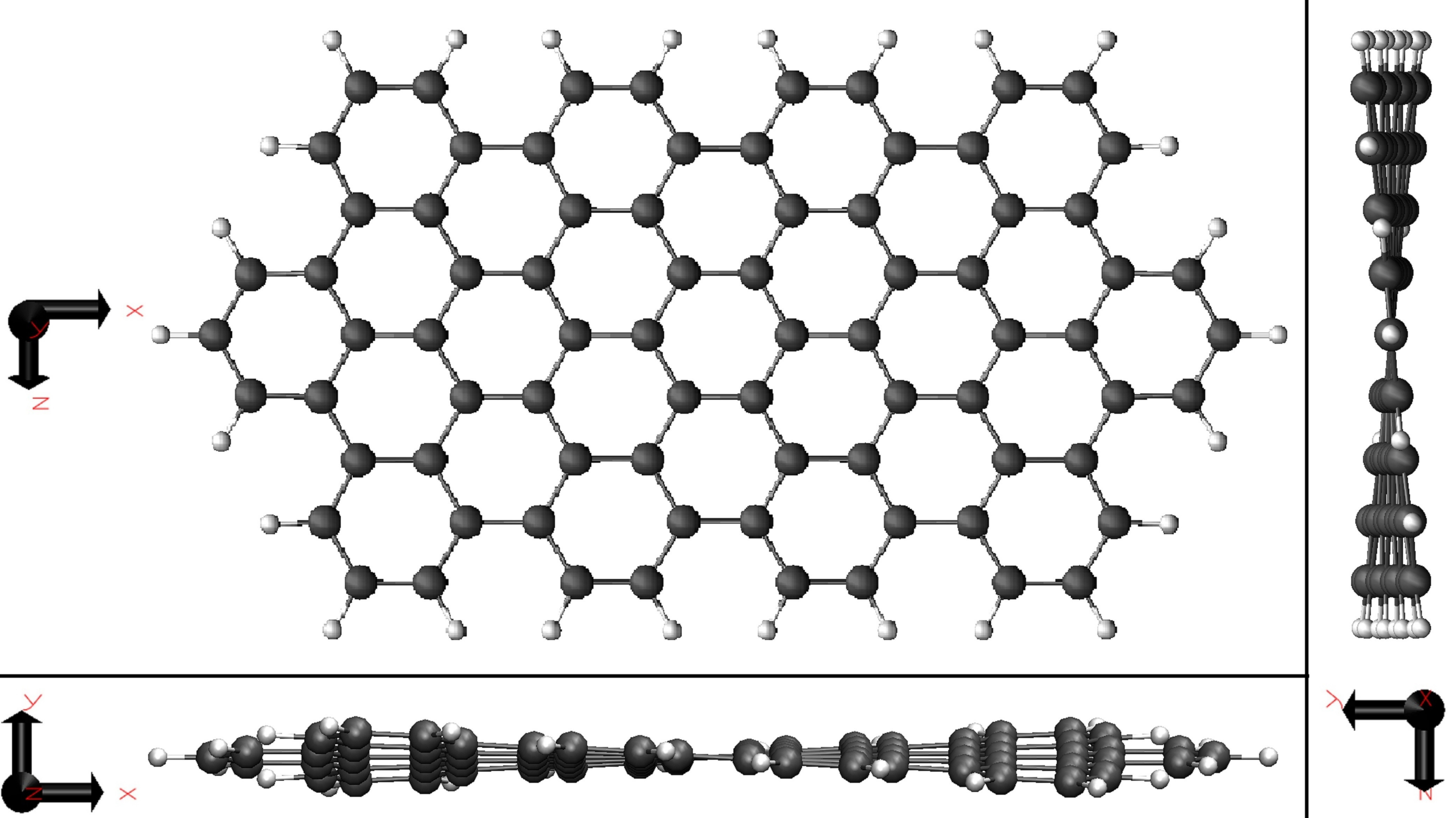}
  \caption{The computed structure of the equilibrium geometry (C$_2$) with the absolute minimum energy of C$_{78}$H$_{26}$ cation is shown, viewed from different angles to better show its 3-D configuration. }
  \label{fig3}
\end{figure}

\begin{figure}[t]
  \centering
  \includegraphics[width=\hsize]{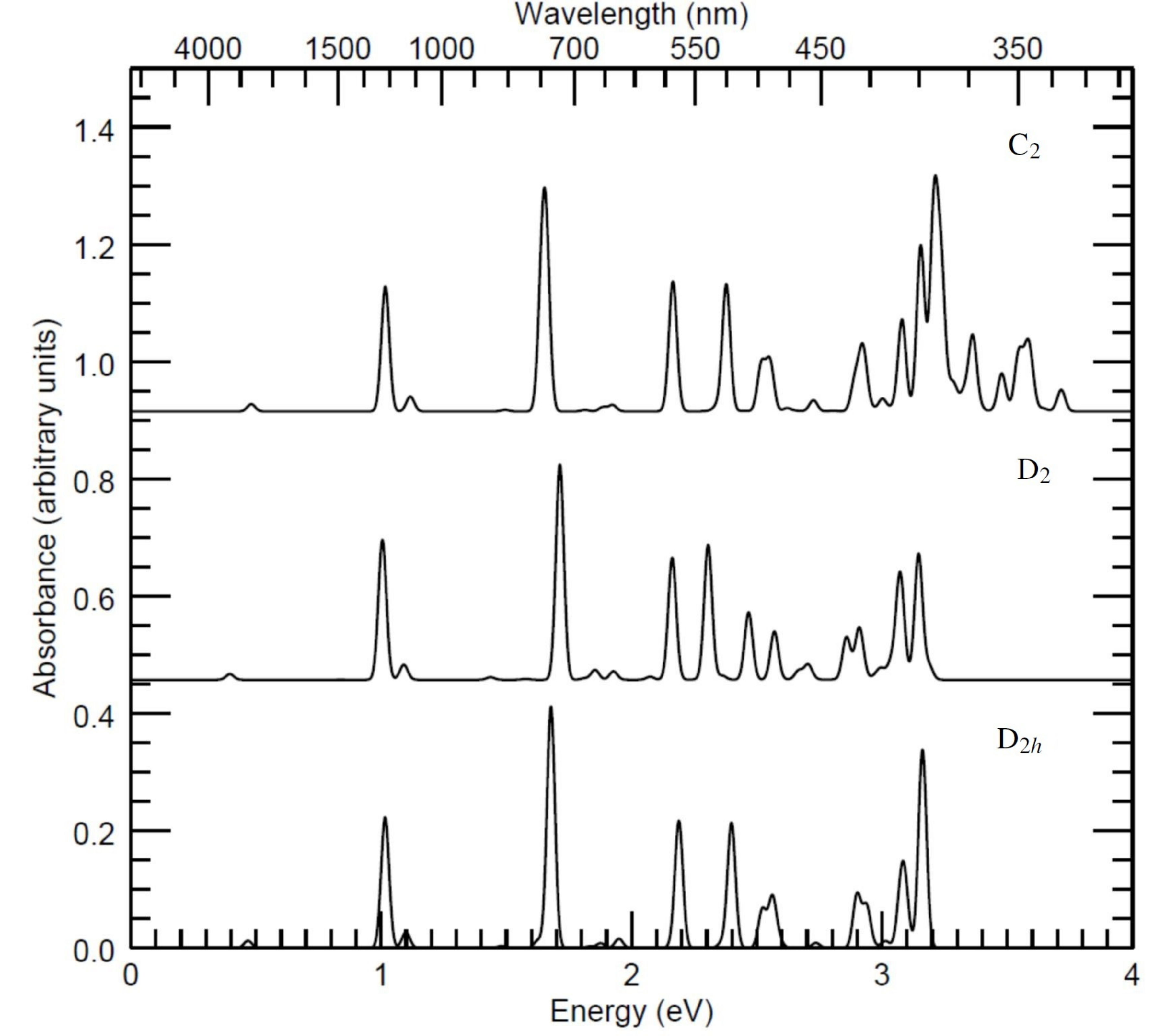}
  \caption{TD-DFT computed spectra at three different molecular geometries, from top to bottom the absolute minimum energy stable configuration (C$_2$ symmetry), another relative minimum (D$_2$ symmetry), and a transition state (D$_{2h}$ symmetry).
  }
  \label{fig4}
\end{figure}

\begin{table*}
\caption{The calculated excited states in nm (eV) with their corresponding oscillator strengths for cationic C$_{78}$H$_{26}$, the equilibrium geometry is C$_2$, D$_2$ and D$_{2h}$. The accuracy of TD-DFT calculations is not better than 0.1 eV, however, we give more digits to distinguish close bands.
\label{1PMCalcs}}\small
\begin{tabular}{ccccccccc}
\hline
\multicolumn{2}{c}{C$_2$}&\multicolumn{2}{c}{D$_2$}&\multicolumn{2}{c}{D$_{2h}$}\\

Excitation Energy/nm(eV)&f-value&Excitation Energy/nm(eV)&f-value&Excitation Energy/nm(eV)&f-value\\

393.08	(3.154) &	0.273&	389.02	(3.187) &	0.022&	390.95	(3.171) &	0.002\\
402.60	(3.079) &	0.155&	394.15	(3.145) &	0.213&	392.22	(3.161) &	0.331\\
407.47	(3.042) &	0.005&	402.17	(3.083) &	0.032&	401.09	(3.091) &	0.094\\
413.06	(3.001) &	0.017&	403.89	(3.069) &	0.154&	403.39	(3.073) &	0.075\\
419.46	(2.955) &	0.003&	408.40	(3.036) &	0.027&	411.40	(3.013) &	0.011\\
423.92	(2.924) &	0.099&	413.83	(2.996) &	0.017&	419.79	(2.953) &	0.003\\
426.63	(2.906) &	0.019&	417.01	(2.973) &	0.005&	422.03	(2.938) &	0.068\\
428.95	(2.890) &	0.044&	425.24	(2.915) &	0.038&	427.47	(2.900) &	0.074\\
454.86	(2.725) &	0.018&	426.80	(2.905) &	0.052&	427.96	(2.897) &	0.016\\
473.10	(2.620) &	0.005&	432.83	(2.864) &	0.013&	478.89	(2.589) &	0.011\\
485.77	(2.552) &	0.083&	434.03	(2.856) &	0.060&	483.92	(2.562) &	0.081\\
492.44	(2.517) &	0.078&	458.40	(2.704) &	0.025&	491.71	(2.521) &	0.065\\
521.41	(2.378) &	0.216&	482.45	(2.570) &	0.082&	516.90	(2.398) &	0.213\\
530.23	(2.338) &	0.008&	502.56	(2.467) &	0.114&	525.33	(2.360) &	0.008\\
572.70	(2.165) &	0.220&	525.08	(2.361) &	0.007&	566.50	(2.188) &	0.216\\
644.41	(1.924) &	0.011&	537.80	(2.305) &	0.229&	636.06	(1.949) &	0.015\\
657.77	(1.885) &	0.007&	573.48	(2.162) &	0.207&	677.32	(1.830) &	0.002\\
747.47	(1.658) &	0.251&	643.32	(1.927) &	0.014&	738.97	(1.677) &	0.411\\
755.22	(1.641) &	0.187&	723.58	(1.713) &	0.366&	762.11	(1.626) &	0.013\\
\hline
\end{tabular}
\end{table*}

\section{Computational methods and results}
\label{sec:com}
Our theoretical calculations were carried out in the framework of the DFT, in its stationary formulation for molecular properties in the electronic ground state, and in its time-dependent formulation (TD-DFT) to study excited states. In particular, we used the B-LYP gradient-corrected exchange-correlation functional \citep{lee88} in combination with the def2-TZVP Gaussian basis set \citep{sch93}, under the resolution of identity approximation, as implemented in the Turbomole \citep{wei05} computer code. This combination of functional and basis set was found to be appropriate and reliable for PAH cations, both for ground electronic state geometry optimization and harmonic vibrational analyses, and for the subsequent calculation of the electronic photo-absorption spectrum via TD-DFT.

To speed up the calculation, a first rough approximation of the equilibrium geometry was obtained making use of the built-in force-field optimization in the Avogadro computer code \citep{han12}. From that starting point, which has D$_{2h}$ symmetry, we obtained the geometry of minimum energy (namely $-2987.9585$ Hartree total DFT energy) at the B-LYP/def2-TZVP theory level, still constrained to D$_{2h}$ symmetry. However, a subsequent harmonic vibrational analysis at the same theory level revealed that geometry to be not a true minimum, but instead a saddle point of the energy, with three imaginary (even if very small) imaginary vibrational frequencies. We therefore perturbed the geometry along the normal coordinates corresponding to the imaginary frequencies, resulting in a lowered symmetry, and restarted the geometry optimization, to find new stationary points. Iterating this procedure, we found an absolute minimum of the energy ($-2988.0585$ Hartree) at a geometry with C$_2$ symmetry, another relative minimum ($-2988.0427$ Hartree) at a different geometry with D$_2$ symmetry, a saddle point ($-2988.0470$ Hartree) with one imaginary frequency at a geometry with C$_{2h}$ symmetry, and another one ($-2987.9252$ Hartree) at a geometry with C$_{2v}$ symmetry. Fig. 3 shows the C$_2$ symmetry geometry viewed from three dimension angles.

We remark that the minimum energy configurations, both the absolute minimum with C$_2$ symmetry and the relative one with D$_2$ symmetry, can be obtained from the more symmetric D$_{2h}$ geometry distorting the molecule in two opposite directions, hence obtaining distinct, different minima separated by potential barriers not higher than $\Delta$E(D$_{2h}$-C$_2$)=2.72eV, $\Delta$E(D$_{2h}$-D$_2$)=2.41eV. A detailed study of the electronic energy hypersurface would be needed to find out whether the different minima can be connected by paths with lower barriers, but this is out of the scope of the present work. This overall configuration, reminiscent of species like corannulene (C$_{20}$H$_{10}$) or NH$_3$, will cause the actual wavefunctions to be superpositions of those centered on each minimum, thereby recovering the full D$_{2h}$ symmetry and yielding groups of very close states, which will in turn produce fine structure in the electronic transitions (i.e. as the inversion doubling in NH$_3$). This structure is completely lost in the broad bands detected by the MPD spectrum, so we do not study it quantitatively here. 

We thereafter performed TD-DFT calculations, still at the B-LYP/def2-TZVP level of theory, to compute the low-energy part of the electronic photo-absorption spectrum of C$_{78}$H$_{26}$$^+$, respectively at the absolute minimum geometry with C$_2$ symmetry, at the relative minimum with C$_2$ symmetry, and at the planar transition state with D$_{2h}$ symmetry. Table~\ref{1PMCalcs} lists the most intense transitions in these geometries, and Fig.~\ref{fig4} shows them in visual form, arbitrarily assuming a FWHM of $\sim$0.04eV.

\section{Discussion}
\label{sec:dis}
Figure~\ref{fig2} compares our MPD spectrum of C$_{78}$H$_{26}$$^+$ , obtained with (7.0 $\pm$ 1.0) mJ per pulse and 2.0 s irradiation time, with the theoretical one computed by TD-DFT for the molecule in its minimum energy geometry (C$_2$ symmetry).
We identify the first band at 431\,nm in the MPD spectrum with the band at 423.9\,nm (f=0.099), possibly merged with the very close ones at 426.6\,nm (f=0.019), and 429.0\,nm (f=0.044) in the TD-DFT vertical spectrum. The second band measured at 516\,nm would correspond to the transition predicted at 521.4\,nm (f=0.216) in the theoretical spectrum. The TD-DFT spectrum in addition predicts bands at 485.8\,nm (f=0.083), at 492.4\,nm (f=0.078), and at 572.7\,nm (f=0.220), which are not detected in the MPD spectrum.

There is a ratio of about 8 between the peak yields of the two MPD bands, compared to a ratio between 1.3 (if one considers the sum of the three close bands near 426nm) and 2.2 (if one considers only the strongest band at 423.9nm) between the theoretical oscillator strengths of the matching transitions. This is because the intensity of the MPD signal is roughly proportional to the probability for the molecule to absorb, sequentially, as many photons as are needed for it to dissociate. Higher energy transitions require fewer photons to dissociate the molecule, so between two transitions having the same oscillator strength but different energy then the one at higher energy will show a stronger MPD signal. If we assume that the dissociation threshold is $\sim$23eV \citep{mon13}, it takes about $\ge$8 photons at 431\,nm (2.88\,eV) to dissociate the molecule. At the MPD peak at about 516nm (2.40\,eV) $\ge$10 photons are required for dissociation, as well as for the undetected bands predicted by TD-DFT at 485.8\,nm (2.55\,eV) and 492.4\,nm (2.52\,eV), but these are about half as intense. TD-DFT calculations also predict a strong band (f=0.220) at 572.7\,nm (2.17\,eV), but it would take $\ge$11 photons to achieve dissociation. If the 23 eV of internal energy that is necessary for dissociation is reached by accumulating energy over several laser pulses, which is likely to be the case, then a critical parameter is the cooling between pulses. In particular at high energy, efficient cooling is expected via electronic emission \citep{berne15, leger88, ander01, martin13}. 
This cooling will delay the heating by multiple photon absorption, making even more inefficient the events that require a larger number of absorbed photons.

Another factor to take into account is that all photons but the first are absorbed by a more and more vibrationally excited molecule: if the energy of a given transition is very sensitive to the precise geometric configuration of the ions, in a simplistic semiclassical description its precise position will oscillate with molecular vibrations, possibly out of the wavelength range of the irradiating laser. A less intuitive but more precise description would be that each band will display a complex vibronic structure, involving absorption from more and more vibrationally excited states. 
To first order, in the Franck\textendash Condon approximation, the total integrated intensity of all vibronic bands will remain constant, but with increasing temperature it will be spread among an increasing number of vibronic bands, leading to a T\textendash dependent broadening and shift of the band. Even if the laser is centered on the absorption peak at 0K, each absorbed photon will heat the molecule, decreasing the absolute absorption cross\textendash section at the laser wavelength. Depending on how sensitive to temperature a specific band is, the absorption cross\textendash section "seen" by the last absorbed photons that leads to dissociation may be very much lower than the peak at 0K. 
This effect has been quantitatively studied for the methyl-pyrene cation by \citet{rapacioli2015}, by coupling ab\textendash initio molecular dynamics with the repeated calculation of vertical electronic spectra. For some of the bands they find that the absolute absorption cross\textendash section at the 0K peak wavelength can decrease by more than an order of magnitude at 500K, due to the combined effect of band shift and broadening. 
If the wavelength range covered by the laser is narrow ($\sim$5~cm$^{-1}$ in our experiment), this may severely reduce the probability to achieve dissociation by the sequential absorption of a large number of photons in a single\textendash color experiment like ours. In addition, the temperature dependence of the spectrum will cause a substantial, artificial band broadening, connected to the experimental technique.
In general, a transition among electronic states described by parallel potential energy surfaces, with almost coincident minima, will be narrow and minimally affected by temperature effects whereas transitions among states with widely different equilibrium geometries will exhibit long vibronic sequences and large variations with temperature.

To have at least an estimate of how important this effect can be for different transitions of the molecule we are studying, we can compare the TD-DFT spectra at some other slightly different geometries, namely the ones corresponding to some of the other stable conformations or transition states.
These are listed in Table~\ref{1PMCalcs}, and visually shown in Fig.~\ref{fig4}. Differences are rather apparent, with some bands being almost unaffected by these specific geometry distortions while others shift quite visibly.

In general, the strongly nonlinear relation between MPD intensities and oscillator strengths, together with the more or less important dependence on vibrational temperature of each given electronic photoabsorption band, produces a strong selection effect on which bands can be detected by MPD, and which not: MPD will typically (relatively) easily detect only the strongest bands, and the ones whose position and profile, at the resolution given by laser line width, are less sensitive to vibrational temperature. A consequence of this latter selection effect is that band positions, for strong bands detected by MPD, are likely to be very close to the ones of the bands measured in absorption at the temperature of the experiment. This side effect makes this technique suitable to test candidate carrier molecules for the astronomical DIBs, whose identification requires very accurate band position measurements. We note that within the region of the spectrum plotted in Fig. 2, there are several DIBs but none precisely matching the bands observed in the experimental spectrum. The closest DIBs to the feature we measured at 431~nm are the ones at 4176~\AA\ \citep{jen94} and 4428~\AA\ \citep{BB37}; the closest to 516~nm is the DIB at 5109~\AA\ \citep{jen94}.

\section{Conclusion}
\label{sec:concl}
We have succeeded in obtaining an optical spectrum of the isolated giant PAH cation C$_{78}$H$_{26}$$^+$. Its strongest transitions in the region studied were found to lie at 431\,nm and 516\,nm, which does not correspond to any known DIB. Despite drawbacks on band profiles and intensities, the MPD technique is expected to provide a reliable band position for absorption bands and then allows us to identify candidates of interest for the DIBs that could motivate spectroscopic studies with more sophisticated techniques as was recently used for the C$_{60}$$^+$ spectrum \citep{camp15}. In particular, our approach can be used to obtain the spectra of still larger and less symmetric systems which may exhibit much stronger transitions. 

This work also motivated the study of C$_{78}$H$_{26}$$^+$ from a theoretical point view. One of the byproducts is the remarkable floppiness of this large PAH and the non\textendash planarity of its lowest energy configuration. We predict several stable geometries close in energy, separated by relatively small barriers. Although the resulting spectroscopic fine structure cannot be observed in the band profile in an MPD experiment, it should be observable in low-temperature absorption spectra such as those producing the DIBs in space.

In addition, the fact that C$_{78}$H$_{26}$$^+$ spontaneously breaks the planar symmetry together with the very low frequencies of the normal modes corresponding to flopping motions, hints that such species can easily fold in closed structures like nanotubes or cages, giving some theoretical support to the finding of \citep{zhen2014} that photoprocessing of C$_{78}$H$_{26}$$^+$ can be a source of fullerenes.

\section{Acknowledgments}
We are grateful to L. Nogu\`es for technical support on the PIRENEA setup and and to L. J. Allamandola for providing the chemical sample from the Ames collection. We acknowledge support from the European Research Council under the European Union's Seventh Framework Programme ERC-2013-SyG, Grant Agreement n. 610256 NANOCOSMOS.


\begin{thebibliography}{00}

\bibitem[Andersen et al. (2001)]{ander01} Andersen, J., Gottrup, C., Hansen, K., Hvelplund, P., \& Larsson, M. 2001, Eur. Phys. J. D, 17, 189.

\bibitem[Allamandola et al.(1989)]{all89}Allamandola, L. J., Tielens, A. G. G. M., Barker, J. R. 1989, Astrophys. J. Suppl. Ser., 71, 733-775.

\bibitem[Beals \& Blanchet (1937)]{BB37} Beals, C. S. and Blanchet, G. H. 1937, PASP, 49, 224.

\bibitem[Bern\'e \&\ Tielens(2012)]{berne12} Bern\'e, O., Tielens, A. G. G. M. 2012, Proc. Natl. Acad. Sci., 109, 401-406.

\bibitem[Bern\'e et al.(2015)]{berne15} Bern\'e, O., Montillaud, J.,  Joblin, C. 2015, Astron. Astrophys., 577, A133.

\bibitem[Boschman et al.(2015)]{bos15} Boschman, L., Spaans, C. M.,  Hoekstra, R., and Schlath{\H o}lter, T. 2015, Astron. Astrophys., 579, A72

\bibitem[Campbell et al. (2015)]{camp15} Campbell1, E. K., Holz1, M., Gerlich, D. and Maier1, J. P. 2015, Nature, 523, 322-323.

\bibitem[Candian et al.(2014)]{can14}Candian, A., Sarre, P.J. and Tielens, A.G.G.M.\ 2014, Astrophys. J. Lett., 791, L10.

\bibitem[Crawford et al. (1985)]{cra85}Crawford, M. K., Tielens, A. G. G. M. and Allamandola, L. J. 1985, Astrophys. J., 293, L45.

\bibitem[Cox(2011)]{cox2011} Cox, N. L. J. 2011, in ``The Diffuse Interstellar Bands'', EAS Pubs. Ser., vol 46, 2011, p. 349.

\bibitem[Genzel et al.(1998)]{gen98}Genzel, R., et al. 1998, Astrophys. J., 498, 579-605.

\bibitem[Hammonds et al.(2009)]{hammonds2009} Hammonds, M., Pathak, A. and Sarre, P.J. 2009, Phys. Chem. Chem. Phys., 11, 4458-4464.

\bibitem[Hanwell et al. (2012)]{han12}Hanwell, M.; Curtis, D. E.; Lonie, D. C.; Vandermeersch, T.; Zurek, E.; Hutchison, G. R. 2012, J. Cheminf., 4, 17.

\bibitem[Huisken et al.(2013)]{huisken2013} Huisken, F., Rouill\'e, G., Steglich, M., Carpentier, Y., J\"ager, C., \& Henning, T. 2013, ``The Diffuse Interstellar Bands'', Proceedings of the International Astronomical Union, IAU Symposium, vol. 297, p. 265

\bibitem[Jennishens \& De\`sert (1994)]{jen94}Jenniskens, P., \& De\`sert, F.-X. 1994, Astron. Astrophys. Ser., 106, 39-78.

\bibitem[Joblin et al. (2002)]{job02}Joblin, C., Pech, C., Boissel, P., Armengaud, M., Frabel, P.  Infrared and Submillimeter Space Astronomy, EAS Pubs. Ser., vol 4, 2002, p. 73.

\bibitem[Kokkin et al.(2013)]{kokkin2013} Kokkin, D. L., Simon, A., Marshall, C., Bonnamy, A., \& Joblin, C.  2013, ``The Diffuse Interstellar Bands'', Proceedings of the International Astronomical Union, IAU Symposium, vol. 297, p. 286

\bibitem[Koskinen et al.(2008)]{kos08}Koskinen, P., Malola, S.\ \&  H\"akkinen, H.\ 2008, Phys. Rev, Lett., 101, 115502(1-4).

\bibitem[Lee et al. (1988)]{lee88}Lee, C., Yang, W., and Parr, R. G. 1988, Phys. Rev. B, 37, 785-789.

\bibitem[Leger et al. (1985)]{leg85}L{\'e}ger, A., \& d\textquoteright Hendecourt, L. 1985, Astron. Astrophys., 146, 81-85.

\bibitem[L\'eger et al. (1988)]{leger88} L\'eger, A., Boissel, P. and d\textquoteright Hendecourt, L. 1988, Phys. Rev. Lett., 60, 921(1-4).

\bibitem[Le Page et al. (2001)]{lep01} Le Page, V., Snow, T. P., Bierbaum, V. M. 2001, Astrophys. J. Suppl. Ser., 132, 233-251.

\bibitem[Marciniak et al.(2015)]{mar15} Marciniak, A., Despre, V., Barillot, T., et al., 2015, Nat. Commun., 6, 7909. 

\bibitem[Martin et al. (2013)]{martin13}Martin, S., Bernard, J., Br\'edy, R.,  Concina, C.,  Joblin,  C.,  Ji, M., Ortega, C., and Chen, L., 2013, Phys. Rev. Lett., 110, 063003 (1-4).

\bibitem[Montillaud et al.(2013)]{mon13}Montillaud, J., Joblin, C., Toublanc, D. 2013, Astron. Astrophys., 552, A15.

\bibitem[M\"uller et al.(1997)]{muller1997} M\"uller, M., Vivekanantan, S.I., K\"ubel, C., Enkelmann, V. and M\"ullen, K. 1997, Angew. Chem. Int. Ed. Engl., 36, 1607

\bibitem[Pino et al.(2011)]{pino2011} Pino, T., Carpentier, Y., F\'eraud, G., Friha, H., Kokkin, D.L., Troy, T.P., Chalyavi, N., Br\'echignac, P., \& Schmidt, T.W.  2011, in ``The Diffuse Interstellar Bands'', EAS Pubs. Ser., vol 46, 2011, p. 355

\bibitem[Poater et al.(2007)]{poa07}Poater, J., Visser, R., Sol\`a,  M.\ \& Bickelhaupt, F.M.\ 2007, J. Org. Chem., 72, 1134-1142.

\bibitem[Puget \&\ Leger (1989)]{pug89}Puget, J. L, Leger, A. 1989, Ann. Rev. Astr. Astrophys., 27, 161-198.

\bibitem[Rapacioli et al.(2015)]{rapacioli2015} Rapacioli, M., Simon, A., Marshall, C.C.M., Cuny, J., Kokkin, D., Spiegelman, F., \& Joblin, C. 2015, accepted for publication in J. Phys. Chem.

\bibitem[Rouill\'e et al.(2013)]{rouille2013} Rouill\'e, G., J\"ager, C., Huisken, F., \&Henning, T. 2013, ``The Diffuse Interstellar Bands'', Proceedings of the International Astronomical Union, IAU Symposium, vol. 297, p. 276

\bibitem[Ruiterkamp(2005)]{ruiterkamp2005} Ruiterkamp, R., Cox, N. L. J., Spaans, M., Kaper, L., Foing, B. H., Salama, F. and Ehrenfreund, P. 2005,  Astron. Astrophys., 432, 515-529.

\bibitem[Salama(1992)]{salama1992} Salama, F. and Allamandola, L. J. 1992, Astrophys. J., 395, 301.

\bibitem[Salama et al. (1999)]{sal99}Salama, F., Galazutdinov, G. A., Krelowski, J., Allamandola, L. J., \& Musaev, F. A. 1999,  Astrophys. J., 526, 265.

\bibitem[Salama \& Ehrenfreund(2013)]{salama2013} Salama, F., \& Ehrenfreund, P. 2013, ``The Diffuse Interstellar Bands'', Proceedings of the International Astronomical Union, IAU Symposium, vol. 297, p. 364

\bibitem[Sarre(2013)]{sarre2013} Sarre, P. J. 2013, ``The Diffuse Interstellar Bands'', Proceedings of the International Astronomical Union, IAU Symposium, vol. 297, p. 34

\bibitem[Schafer et al. (1993)]{sch93}Schafer, A., Huber, C., and Ahlrichs, R. 1993, J. Chem. Phys., 100, 5829-5835.

\bibitem[Sellgren(1984)]{sel84}Sellgren, K. 1984,  Astrophys. J., 277, 623.

\bibitem[Tan(2009)]{tan2009} Tan, X. 2009, Spectrochimica Acta A, 71, 2005-2011.

\bibitem[Tielens(2013)]{tie13}Tielens, A. G. G .M.\ 2013, Rev. Mod. Phys., 85, 1021.

\bibitem[Tielens(2008)]{tie08}Tielens, A. G. G. M.\ 2008, Ann. Rev. Astr. Astrophys., 46, 289-337.

\bibitem[Useli et al. (2010)]{use10} Useli-Bacchitta, F., Bonnamy, A., Malloci, G., Mulas, G., Toublanc, D., Joblin, C. 2010, Chem. Phys. 371, 16-23.

\bibitem[Walker et al. (2015)]{wal15} Walker, G. A. H., Bohlender, D. A., Maier, J. P.,  Campbell, E. K. 2015, Astrophys. J. Lett., 812, L8.

\bibitem[Weigend \& ahlrichs (2005)]{wei05}Weigend, F., and Ahlrichs, R. 2005, Phys. Chem. Chem. Phys., 7, 3297-3305.

\bibitem[Weisman et al.(2003)]{weisman2003} Weisman, J. L., Lee, T. J., Salama, F. and Head\textendash Gordon, M. 2003, Astrophys. J., 587, 256-261.

\bibitem[West et al. (2014)]{wes14}West, B., Useli-Bacchitta, F., Sabbah, H., Blanchet, V., Bodi, A., Mayer, P. M. and Joblin, C. 2014, J. Phys. Chem. A, 118, 7824-7831.

\bibitem[Zhen et al.(2014)]{zhen2014} Zhen, J., Castellanos, P., Paardekooper, D.~M., Linnartz, H., \ Tielens, A.~G.~G.~M.\ 2014, Astrophys. J. Lett., 797, L30.

\end{thebibliography}
\end{document}